\documentclass[12pt]{amsart}
\usepackage{amsmath,amssymb,amsthm,amsfonts,stmaryrd,graphicx}
\usepackage{stmaryrd}
\usepackage[english]{babel}
\usepackage{caption}
\newtheorem{thm}{{\bf Theorem}}[section]

\newtheorem{lem}{{\bf Lemma}}[section]
\newtheorem{prop}{{\bf Proposition}}[section]

\theoremstyle{definition}

\theoremstyle{remark}
\newtheorem{rem}{{\bf Remark}}[section]
\numberwithin{equation}{section}

\newcommand{\R}{\mathbb{R}}
\newcommand{\N}{\mathbb{N}}

\newcommand{\C}{\mathbb{C}}

\newcommand{\ds}{\displaystyle}

\everymath{\displaystyle}
\newcommand{\ef}{\quad \hfill$\blacksquare$\vspace{0.15cm}\\}

\begin{document}

\title[Generalized discrete $q$-Hermite II Polynomials \& $q$-deformed oscillator]{ Generalized discrete $q$-Hermite II Polynomials and $q$-deformed oscillator}%
\author{  Kamel Mezlini }%
\address{ Kamel  Mezlini.University of Carthage,
High Institute of Applied Sciences and Technologies of Mateur, 7030, Tunisia.}
\email{kamel.mezlini@lamsin.rnu.tn}%
\email{kamel.mezlini@yahoo.fr}%

\begin{abstract}
In this paper, we present an explicit  realization of $q$-deformed Calogero-Vasiliev algebra  whose generators are first-order $q$-difference operators  related to the  generalized discrete $q$-Hermite II polynomials recently introduced
in \cite{ghermite}. Furthermore, we construct the wave functions and we determine the  $q$-coherent states.

\end{abstract}
\maketitle
\section{Introduction}

The   $q$-deformed harmonic oscillator  algebras  \cite{ Drinfel,Floreanini,Kulish, Macfarlane1}  have been intensively studied in
recent years due to their crucial role in diverse areas of mathematic and physics.
The basic interest in $q$-deformed algebras resides in the generalization of the fundamental symmetry concept of
classical  Lie algebras.

Many  algebraic constructions have been proposed  to describe various  generalization of the quantum harmonic oscillator in the literature.
The difficulty for most of them is to realize an explicit form of the associated Hamiltonian eigenfunctions.
It is well known that the  Hermite  polynomials are connected to the realization of the classical
harmonic oscillator algebra. It is natural then, that generalizations of quantum harmonic oscillator  lead to generalizations of the Hermite polynomials.
An explicit realization of the $q$-harmonic oscillator has
has been explored by many  authors see for example
Atakishiev \cite{Atakishiyev, Atakishiyev2}, Borzov \cite{Borzov}, also Kulish and Damaskinsky \cite{Kulish},
where the eigenfunctions of the
corresponding  Hamiltonian are given explicitly in terms of the
$q$-deformed Hermite polynomials.
The generators of the corresponding algebra are realized in terms of first-order difference
operators.\\
In particular,  as pointed out by Macfarlane in \cite{Macfarlane1,Macfarlane}, the Calogero-Vasiliev oscillator  generalizes the  parabose oscillator and its $q$-deformation  describes the $q$-analogue of the parabose oscillator \cite{Floreanini}.
In one dimensional case, Rosenblum in \cite{Rosenblum} studied the generalized Hermite polynomials associated with the Dunkl operator and used them to construct the eigenfunctions of the parabose ossillator Hamiltonian . This oscillator, as it has been shown in \cite{Macfarlane}, is linked to two-particle Calogero model \cite{Brink}.\\
The purpose of this paper is to explore the generalized discrete $q$-Hermite II polynomials $ \tilde h_{n,\alpha}(x;q)$, recently introduced in  \cite{ghermite} to construct the  Hamiltonian eigenfunctions for  the $q$-deformed Calogero-Vasiliev oscillator.
This allows to find an explicit form  of the generators of the corresponding algebra in terms of $q$-difference operators.\\
This paper is organized as follows: in Section 2, we recall some notations and useful results
 from \cite{ghermite} about the  generalized discrete $q$-Hermite II polynomials $ \tilde h_{n,\alpha}(x;q)$.
In Section 3, We review briefly the Fock space description of the Calogero-Vasiliev oscillator and its $q$-deformation as developed by  Macfarlane in \cite{Macfarlane1,Macfarlane}.
In Section 4, we introduce an explicit form of the eigenfunctions of  the $q$-deformed Calogero-Vasiliev Hamiltonian oscillator. This directly leads to the dynamical symmetry algebra $su_{q^\frac{1}{2}}(1, 1)$, whose
generators are explicitly constructed in terms of the $q$-difference operators,
we construct the family of coherent states of this oscillator. Finally,  we investigate the limiting case of the $q$-deformed Calogero-Vasiliev oscillator.

\section{Notations and Preliminary}
For the convenience of the reader, we provide in this section a summary of the
mathematical notations and definitions used in this paper.
We refer to the
general references \cite{G}  and \cite{ghermite} for the definitions and notations. Throughout this paper, we assume that $0<q<1$ and
we write
 $\mathbb{R}_{q}=\{\pm q^{n}, n \in \mathbb{Z}\}$.
\subsection{Basic symbols}
For complex a number $a$, the $q$-shifted factorials are defined by:
$$(a;q)_{0}=1;\ \ \ (a;q)_{n}=\prod_{k=0}^{n-1}(1-aq^{k}),
n=1, 2,...;\ \ \ (a;q)_{\infty}=\prod_{k=0}^{\infty}(1-aq^{k}).$$
The $q$-numbers and the $q$-factorials are defined as follows:
\begin {equation*}
\llbracket  x\rrbracket_{q}={{1-q^x}\over{1-q}}, ~~~~~ x\in \C\quad {\rm and}\quad n!_q =\llbracket  1\rrbracket_{q}\llbracket  2\rrbracket_{q}...\llbracket  n\rrbracket_{q},   ~~~~~~\llbracket  0\rrbracket_{q}=1,  ~~~~~~n\in \N.
\end {equation*}
For $\alpha\in\R$, we define the generalized $q$-integers and the generalized $q$-factorials  by
\begin{equation}\label{nfactalpha}
\left\llbracket2n \right\rrbracket_{q,\alpha}=\left\llbracket 2n\right\rrbracket_{q},\;\;\;\left\llbracket 2n+1\right\rrbracket_{q,\alpha}=\left \llbracket2n+2\alpha+2\right\rrbracket_{q};\;\;
n!_{q,\alpha}=\left\llbracket1 \right\rrbracket_{q,\alpha}\left\llbracket2 \right\rrbracket_{q,\alpha}...\left\llbracket n \right\rrbracket_{q,\alpha}
\end{equation}
and the generalized $q$-shifted factorials by
\begin{equation}\label{qqnalpha}
(q;q)_{n,\alpha}:=(1-q)^nn!_{q,\alpha}.
\end{equation}
Remark that we can  rewrite (\ref{qqnalpha}) as
\begin{equation*}
\begin{array}{lll}
(q;q)_{2n,\alpha}&=&(q^2;q^2)_{n}(q^{2\alpha+2};q^2)_{n},\\
(q;q)_{2n+1,\alpha}&=&(q^2;q^2)_{n}(q^{2\alpha+2};q^2)_{n+1}.
\end{array}
\end{equation*}
We may express the generalized $q$-factorials as
\begin{equation*}
\begin{array}{lllll}
(2n)!_{q,\alpha}&=& \ds\frac{(1+q)^{2n}\Gamma_{q^2}(\alpha+n+1)\Gamma_{q^2}(n+1)}{\Gamma_{q^2}(\alpha+1)},\\
(2n+1)!_{q,\alpha}&=& \ds\frac{(1+q)^{2n+1}\Gamma_{q^2}(\alpha+n+2)\Gamma_{q^2}(n+1)}{\Gamma_{q^2}(\alpha+1)},
\end{array}
\end{equation*}
where $\Gamma _q$ is the $q$-Gamma function  given by (see \cite{Sole,G} )
\begin{equation*}
\Gamma _q(z)=\frac {(q;q)_\infty }{(q^z;q)_\infty }(1-q)^{1-z}, \
z\neq 0,-1,-2,\dots \label{1.57nn}
\end{equation*}
and tends to $\Gamma (z) $ when $q$ tends to $1^{-}$.
In particular, we have the limits
\begin{equation}\label{limitefac}
\begin{array}{lll}
\ds\lim_{q\rightarrow1^-}(2n)!_{q,\alpha}&=& \ds\frac{2^{2n}n!\Gamma(\alpha+n+1)}{\Gamma(\alpha+1)}=\gamma_{\alpha+\frac{1}{2}}(2n), \\
\ds\lim_{q\rightarrow1^-}(2n+1)!_{q,\alpha}&=&  \ds\frac{2^{2n+1}n!\Gamma(\alpha+n+2)}{\Gamma(\alpha+1)}=\gamma_{\alpha+\frac{1}{2}}(2n+1),
\end{array}
\end{equation}
where $\gamma_{\nu}$ is  the Rosenblum's generalized factorials (see \cite{Rosenblum}).

\begin{rem}\label{relaqqalpha}
If $\alpha=-\frac{1}{2}$, then we get
$(q;q)_{n,\alpha}= (q;q)_{n}\;\;\mbox{and}\;\;n!_{q,\alpha }=n!_{q}$.
\end{rem}

\subsection{The generalized $q$-exponential functions}
\noindent The two Euler's   $q$-analogues of the  exponential
function are given by ( see \cite{G})
\begin{equation}\label{q-Expo}
E_q(z)
=\displaystyle\sum_{k=0}^{\infty}\displaystyle\frac{q^{\frac{k(k-1)}{2}}z^{k}}{(q;q)_{k}}=(-z;q)_{\infty}
\end{equation}
and
\begin{equation}\label{q-expo}
e_q(z)
=\displaystyle\sum_{k=0}^{\infty}\displaystyle\frac{z^{k}}{(q;q)_{k}}=\displaystyle\frac{1}{(z;q)_{\infty}},\;\;\;\;|z|<1.
\end{equation}

\noindent For $z\in\C$, the generalized $q$-exponential functions are defined by ( see \cite{ghermite})
\begin{equation}\label{q-Expo-alpha}
E_{q,\alpha}(z):=\displaystyle\sum_{k=0}^{\infty}\displaystyle\frac{q^{\frac{k(k-1)}{2}}z^{k}}{(q;q)_{k,\alpha}},
\end{equation}
\begin{equation}\label{q-expo-alpha}
e_{q,\alpha}(z):=\displaystyle\sum_{k=0}^{\infty}\displaystyle\frac{z^{k}}{(q;q)_{k,\alpha}},\;\;\;\;|z|<1
\end{equation}
and
\begin{equation}\label{exprubalpha}
\psi_\lambda^{\alpha,q}(z)= \ds\sum_{n=0}^\infty b_{n,\alpha}(i\lambda z;q^2),\;\;\lambda\in\C,
\end{equation}
where
\begin{equation}\label{bn-alpha-def}
b_{n,\alpha}(z;q^2)=\ds\frac{q^{[\frac{n}{2}]([\frac{n}{2}]+1)}z^n}{n!_{q,\alpha}}
\end{equation}
 and $[x]$ denoting the integer part of $x\in\mathbb{ R}$.
 Note that  $\psi_\lambda^{\alpha,q}(z)$ is  the  $q$-Dunkl kernel defined in \cite{Neji_Rym}.\\
A particular case, where $\alpha=-\frac{1}{2}$, by Remark \ref{relaqqalpha} it follows that $E_{q,\alpha}(z) = E_{q}(z)$ and $e_{q,\alpha}(z)=e_{q}(z) $.

\subsection{The generalized $q$-derivatives }
\noindent The Jackson's $q$-derivative  $D_{q}$  (see \cite{G,KC})
is defined by :
\begin{equation}\label{jacksonD}
 D_{q}f(z)=   \frac{f(z)-f(qz)}{(1-q)z}.
\end{equation}
\noindent We also need a variant $D_q^+$, called forward
$q$-derivative of the (backward) $q$-derivative $D_q^- = D_q$ as
defined in (\ref{jacksonD}):
\begin{equation}\label{jacksonD+}
 D_{q}^+f(z)=   \frac{f(q^{-1}z)-f(z)}{(1-q)z}.
\end{equation}
\noindent Note that $  \lim _{q\rightarrow 1^- }D_qf(z)=\lim
_{q\rightarrow 1^- }D_q^+f(z) =f^{\prime}(z)$
whenever $f$ is differentiable at $z$.\\

The generalized backward and forward $q$-derivative operators $D_{q,\alpha}$ and  $D_{q,\alpha}^+ $ are defined as ( see \cite{ghermite})
\begin{equation}\label{D_qalpha}
 D_{q,\alpha}f(z)=   \frac{f(z)-q^{2\alpha+1}f(qz)}{(1-q)z},
\end{equation}
\begin{equation}\label{D_qalphap}
 D_{q,\alpha}^+f(z)=  \frac{f(q^{-1}z)-q^{2\alpha+1}f(z)}{(1-q)z}.
\end{equation}
The  generalized $q$-derivatives operators are given by
\begin{equation}\label{defdeltaalpha}
\Delta _{\alpha,  q}f =D_{q}f_e+D_{q,\alpha}f_o,
\end{equation}
\begin{equation}\label{defdeltaalphaplus}
\Delta _{\alpha,  q}^+f =D_{q}^+f_e+D_{q,\alpha}^+f_o,
\end{equation}
where $f_e$ and $f_o$ are respectively the even and the odd parts of $f$.\\
For $\alpha=-\frac{1}{2}$, we have  $\;D_{q,\alpha}= D_{q}$, $\;\;D_{q,\alpha}^+= D_{q}^+$, $\;\;\Delta_{q,\alpha}= D_{q}\;$  and $\;\Delta_{q,\alpha}^+= D_{q}^+$.\\
We can rewrite the $q$-Dunkl operator introduced in \cite{Neji_Rym} by means of the  generalized $q$-derivative operators as
\begin{equation}\label{dunklopera2}
\Lambda _{\alpha,  q}f= \Delta _{\alpha,  q}^+f_e+ \Delta _{\alpha,  q}f_o.
\end{equation}
\noindent It is noteworthy that
for a differentiable function $f$, we have
\begin{equation}\label{limdunkl}
\lim_{q\rightarrow 1} \Delta _{\alpha,  q}f=\lim_{q\rightarrow 1} \Delta _{\alpha,  q}^+f= \Lambda _{\alpha+\frac{1}{2}}f,
\end{equation}
where $\Lambda_\nu $ is the classical Dunkl operator defined by
\begin{equation}\label{dunkl}
\Lambda_\nu f(x)= f^{\prime}(x)+\frac{\nu}{x}\left[f(x)-f(-x)\right].
\end{equation}
\subsection{The $q$-Dunkl transform}
 We shall  need the Jackson $q$-integral  defined by (see
\cite {G,KC}).
\begin{equation*}\label{eq:integral}
\int_{-\infty}^{\infty}{f(x)d_qx}
=(1-q)\sum_{n=-\infty}^{\infty}q^n f(q^n)+
(1-q)\sum_{n=-\infty}^{\infty}q^nf(-q^n).
\end{equation*}
For $p\geq 1$, we denote by $\ds L_{\alpha ,q}^p(\R_q)$ the space of complex-valued
functions $f$ on $\R_q$ such that
$$\|f\|_{q,p}=\left(\int_{-\infty}^{\infty}|f(x)|^p|x|^{2\alpha +1}d_qx\right)^{\frac{1}{p}}<\infty. $$
The generalized $q$-exponential function  $\psi_\lambda^{\alpha,q}(x)$ defined in (\ref{exprubalpha}) gives rise to a $q$-integral transform, called
the $q$-Dunkl transform on the real line, which was introduced  in \cite{Neji_Rym} as
\begin{equation*}\label{FD}
    \mathcal{F}_D^{\alpha, q}(f)(\lambda) = K_{\alpha}\ds\int_{-\infty}^{+\infty}
    f(x)\psi_{-\lambda}^{\alpha, q}(x) |x|^{2\alpha +1} d_qx,\quad f\in L_{\alpha, q}^1(\R_q),
\end{equation*}
where
$$K_{\alpha}=\ds\frac{(1-q)^\alpha\left(q^{2\alpha+2};q^2\right)_\infty}{2\left(q^{2};q^2\right)_\infty}.$$
It satisfies the following Plancheral theorem : \begin{thm}\label{Plancheral} $ \mathcal{F}_D^{\alpha, q}$ is an isometric isomorphism of   $ L_{\alpha, q}^2(\R_q)$  and  for $f\in L_{\alpha, q}^2(\R_q)$, we have
\begin{equation}
 \| \mathcal{F}_D^{\alpha, q}(f) \|_{q,2} = \| f \|_{q,2}
\end{equation}
\end{thm}
and the following inversion formula
\begin{equation}\label{Inversion}
    f(x) =  K_{\alpha}\ds\int_{-\infty}^{+\infty}
    \mathcal{F}_D^{\alpha,  q}f(\lambda) \psi_x^{\alpha,  q}(\lambda)|\lambda|^{2\alpha +1}
    d_q\lambda,\;\;\forall x\in\R_q.
\end{equation}

\subsection{The generalized discrete $q$-Hermite II polynomials }
The generalized discrete $q$-Hermite II polynomials $\{\tilde h_{n,\alpha}(x;q)\}_{n=0}^\infty$ are defined by (see \cite{ghermite})
\begin{equation}\label{hnalphatild}
\tilde h_{n,\alpha}(x;q):=(q;q)_{n}\ds\sum_{k=0}^{[\frac{n}{2}]}\ds\frac{(-1)^kq^{-2nk}q^{k(2k+1)}x^{n-2k}}{(q^2;q^2)_{k}(q;q)_{n-2k,\alpha}}.
\end{equation}

They  have the following  properties: \\
\begin{itemize}
  \item
\textbf{Generating function}:
\begin{equation}\label{gghntild}
e_{q^2}(-z^2)E_{q,\alpha}( xz)=\ds\sum_{n=0}^\infty\frac{q^{\frac{n(n-1)}{2}}}{(q;q)_n}\tilde  h_{n,\alpha}(x;q)z^n.
\end{equation}
\item \textbf{Inversion formula}:
\begin{equation}\label{monometild}
x^n=(q;q)_{n,\alpha}\ds\sum_{k=0}^{[\frac{n}{2}]}\frac{q^{-2nk+3k^2}\tilde  h_{n-2k,\alpha}(x;q)}{(q^2;q^2)_{k}(q;q)_{n-2k}  }.
\end{equation}
\item \textbf{Forward shift operator}:
\begin{equation}\label{forwardshifttild}
\tilde h_{n,\alpha}(q^{-1}x;q)-q^{(2\alpha+1)\theta_{n+1}}\tilde h_{n,\alpha}(x;q)=q^{-n}(1-q^n)x\tilde h_{n-1,\alpha}(x;q),
\end{equation}
where $\theta_{n}$ is defined to be $0$ if $n$ is odd and $1$ if $n$ is even.\\
\item \textbf{Backward shift operator}:
\begin{equation}\label{backward}
\tilde h_{n,\alpha}(x;q) -q^{(2\alpha+1)\theta_{n+1}}(1+q^{-2\alpha-1}x^2)\tilde h_{n,\alpha}(qx;q)=-q^{n}\frac{ 1-q^{-n-1-(2\alpha+1)\theta_{n}}}{1-q^{-n-1}} x\tilde h_{n+1,\alpha}(x;q).
\end{equation}
 \item\textbf{Orthogonality relation}:
\begin{equation}\label{orthohnalpht}
\ds\int_{-\infty}^{\infty}\tilde h_{n,\alpha}(x;q) \tilde h_{m,\alpha}(x;q) \omega_{\alpha}(x;q)|x|^{2\alpha+1}d_qx= d_{n,\alpha}^{-2}\delta_{n,m},
\end{equation}
where
\begin{equation}\label{omega}
\omega_\alpha(x;q)=e_{q^2}(-q^{-2\alpha-1}x^2)
\end{equation}
and
\begin{equation}\label{dncoef}
d_{n,\alpha}=c_{\alpha}q^{\frac{n^2}{2}}\frac{\left(q;q\right)_{n,\alpha}^{\frac{1}{2}}}{\left(q;q\right)_{n}},\;\;\;
c_{\alpha}=\ds\sqrt{\frac{(-q^{-2\alpha-1}, -q^{2\alpha+3},q^{2\alpha+2};q^2)_\infty}{2(1-q)(-q, -q,q^2;q^2)_\infty }}.
\end{equation}
\end{itemize}
\section{The Calogero-Vasiliev Oscillator and $q$-deformation}

\subsection{The Calogero-Vasiliev Oscillator}
The Calogero-Vasiliev oscillator algebra \cite{Macfarlane1,Macfarlane} (  also called the deformed Heisenberg algebra with reflection   \cite{Brink2})  is generated by the operators $\{I, a, a^+,N,K\}$ subject to the Hermiticity
conditions
\begin{equation}\label{conjug}
 (a^+)^*=a,\;\;N^*=N,\;\; K^*=K^{-1}
\end{equation}
and it satisfies the relations
\begin{equation}\label{calegero}
\begin{array}{llll}
[a,a^+]=I+2\nu K,& \nu\in\R,& 2\nu+1>0, & K^2=I,
\end{array}
\end{equation}
\begin{equation}\label{calegero2}
\begin{array}{lll}
[N,a]=-a, & [N,a^+]=a^+, & [N,K]=0,
\end{array}
\end{equation}
where $[A,B]=AB-BA$.
The operators $a^-$, $a^+$ and $N$  generalize the annihilation, creation
and number operators related to the classical harmonic oscillator.\\
This oscillator, as it has been shown by Macfarlane in \cite{Macfarlane1}, also describes a parabose oscillator of order $p = 2\nu +1$.
In particular, it is linked to two-particle Calogero model \cite{Macfarlane} and  Bose-like oscillator  \cite{Rosenblum}.
This algebra has a basic one-dimensional explicit realization in terms of difference-differential operators
\begin{equation}\label{annih-crea}
A_\nu= \frac{1}{\sqrt{2}}(\Lambda_{\nu} + xI),\;\;\; A_\nu^+= \frac{1}{\sqrt{2}}(\Lambda_{\nu} - xI),
\end{equation}
where  $I$ is the identity mapping and $\Lambda_\nu$ is the Dunkl operator defined by (\ref{dunkl}).
The Hamiltonian is expressed as
\begin{equation*}
H= \frac{1}{2}\left\{A_\nu,A_\nu^+\right\}= \frac{1}{2}(-\Lambda_{\nu}^2+x^2I),
\end{equation*}
 where $\left\{A,B\right\}=AB+BA $.
The eigenvalues of  $H$  are  $n+\frac{1}{2}+\nu$ and the corresponding eigenvectors $\phi_{n}^{\nu}(x)$, which are the generalized Hermite functions introduced by Rosenblum in \cite{Rosenblum} as
\begin{equation}\label{defphimu}
\phi_{n}^{\nu}(x)=\left(\frac{\gamma_{\nu}(n)}{\Gamma(\nu+\frac{1}{2})}\right)^{\frac{1}{2}}exp(-\frac{x^2}{2})\frac{ H_n^{\nu}(x)}{2^{\frac{n}{2}}n!},
\end{equation}
where $\gamma_{\nu}$ is the generalized factorial
\begin{equation*}
\gamma_{\nu}(n)=\frac{2^n\left[ \frac{n}{2}\right]!\Gamma\left(\nu + \left[ \frac{n+1}{2}\right]+\frac{1}{2} \right)}{\Gamma\left(\nu+\frac{1}{2} \right)}
\end{equation*}
and $H_n^{\nu}(x)$ is the generalized Hermite polynomials.
\begin{equation*}
H_n^{\nu}(x):=n!\ds\sum_{k=0}^{[\frac{n}{2}]}\ds\frac{(-1)^k(2x)^{n-2k}}{k!\gamma_\nu(n-2k)}.
\end{equation*}
$\{ \phi_{n}^{\nu}(x)\}_{n\in\N}$ is a complete  orthonormal set in  the  Hilbert space $ L_\nu^2(\R) $ of Lebesgue measurable functions $f$ on $\R$ with
$$ ||f||_\nu:= \left(\int_{-\infty}^{\infty}|f(x)|^2|x|^{2\nu}dx\right)^{\frac{1}{2}}<\infty,$$
on which the conjugation relations (\ref{conjug}) are satisfied.
Let   $ \mathfrak{S}_\nu$ be the  space
spanned by the generalized Hermite functions $\{ \phi_{n}^{\nu}(x)\}_{n=0}^\infty$.
The operators $A_\nu $, $A_\nu^+$ and $N$ act on $\mathfrak{S}_\nu$ as follows
\begin{equation}\label{actionaa+N}
\begin{array}{lll}
A_\nu^+\phi_{2n}^{\nu}(x)&=&\sqrt{2n+2\nu+1}\phi_{2n+1}^{\nu}(x),\\
A_\nu^+\phi_{2n+1}^{\nu}(x)&=&\sqrt{2n+2}\phi_{2n+2}^{\nu}(x),\\
N\phi_{n}^{\nu}(x)&=&n\phi_{n}^{\nu}(x).
\end{array}
\end{equation}
The number operator $N$ is given explicitly in terms of
the creation and annihilation operators by
\begin{equation*}
N=\frac{1}{2}\left\{A_\nu,A_\nu^+\right\}-\frac{2\nu+1}{2}.
\end{equation*}
$K$ is realized in terms of the $N$ operator $K=(-1)^N$.
Obviously, the operators $A_\nu$,  $A_\nu^+$, $N$ and $K$ satisfy the commutation relations (\ref{calegero}) and (\ref{calegero2}) on $\mathfrak{S}_\nu$ .\\
It is well known  that in one dimension the   two-particle Calogero system
realizes  an irreducible representations   of Lie algebra   $su(1, 1)$  \cite{Perelemov}.
Then  one can  easily  verify that the operators
$$K_+=\frac{1}{2}(A_\nu^+)^2,\;\;K_-=-\frac{1}{2}A_\nu^2,\; \hbox{and}\; K_0=\{A_\nu,A_\nu^+\}/4 $$
satisfy the commutation relations
\begin{equation*}
[K_+,K_-]=-2K_0,\;\;[K_0,K_\pm]=\pm K_\pm,\;\;\;\mbox{on}\;\;\mathfrak{S}_\nu.
\end{equation*}
Thus, $K_0$, $K_+$ and $K_-$ are the generators of  Lie algebra $su(1, 1)$.
The representations are characterized
by eigenvalues of the Casimir operator given by
$$C=K_0^2+ \{K_+,K_-\}/2 ,$$
which commutes with $K_0$ and $K_{\pm}$. It follows from  (\ref{actionaa+N})  that $C$ takes the value
$$  -3/16+\nu(\nu\pm1)  $$
throughout the even and odd subspaces of  $\mathfrak{S}_\nu$.
Thus $\mathfrak{S}_{\nu}^\pm$ carry out unitary irreducible representations of $su(1, 1)$ with distinct eigenvalues of the Casimir operator $C$.
\subsection{The $q$-deformed Calogero-Vasiliev oscillator}
The $q$-deformed Calogero-Vasiliev oscillator algebra is defined as the associative initial  algebra generated
by the operators $\{ b,\; b^+,\;N\}$,  which satisfy the relations
\begin{equation}\label{calegero2b}
\begin{array}{lllll}
[N,b]=-b, & [N,b^+]=b^+, &(b^+)^*=b, & N^*=N,
\end{array}
\end{equation}
\begin{equation}\label{commut}
bb^+ - q^{\pm (1+2\nu K)}b^+b= \left[1+2\nu K \right]_{q}q^{\mp (N+\nu -\nu K)},
\end{equation}
where $[x]_q={{q^{x}-q^{-x}}\over{q^{}-q^{-1}}}$ is an alternative definition of $q$-numbers and  $K=(-1)^N$ .\\

The Fock representation of this $q$-oscillator algebra is constructed on  a Hilbert space   $\mathfrak{H}$ with the orthonormal basis  $\{e_n\}_{n=0}^\infty$.
The operators $b,\;b^+$, and $N$   act on  the  subspace $\mathfrak{S}_{q\nu}$ spanned by the basis vectors  $e_n$  according to the formulas (see \cite{Macfarlane1,Macfarlane,Klimyk})
\begin{equation}\label{b+actions}
\begin{array}{llll}
b^+e_{2n}&=& \sqrt{\left[ 2n+2\nu+1\right]_{q}}e_{2n+1},&n=0,1,2,...,\\
b^+e_{2n-1}&=& \sqrt{\left[ 2n\right]_{q}}e_{2n},&n=1,2,...,\\
 Ne_{n} &=&  ne_{n},&n=0,1,2,....\\
\end{array}
\end{equation}
It follows from ( \ref{b+actions})  that  we have the following equalities
\begin{equation}\label{sombb}
bb^+= \left[  N+1+\nu(1+K)\right]_{q}, \;\;\;
b^+b= \left[  N+\nu(1-K)\right]_{q}\;\;\mbox{on}\;\;\mathfrak{S}_{q\nu}.
\end{equation}
The operators $b$, $b^+$ and $N$
directly lead to the realisation of the quantum algebra $su_q(1, 1)$ with the generators (see \cite{Kulish,Macfarlane1,Macfarlane})
$$K_+=\beta(b^+)^2,\;\;\; K_-=\beta b^2,\;\;\;K_0=\frac{1}{2}(N+\nu+\frac{1}{2}),\;\;\;\;\beta=\left([2]_q\right)^{-1}.$$
They satisfy the  commutation relations
$$[K_0,K_{\pm}]=\pm K_{\pm},\;\;\;\;[K_{-}, K_{+}]= \left[ 2K_0 \right]_{q^2}\;\;\mbox{on}\;\;\mathfrak{S}_{q\nu}$$
and the conjugation relations
$$(K_0)^*=K_0,\;\;(K_+)^*=K_-. $$
The Casimir operator $C$, which by definition commutes with the generators $K_{\pm} $ and $K_0$     is
 $$C= \left[ K_0-\frac{1}{2}\right]_{q^2}^2-K_{+}K_{-}.    $$
The action of the  operator $C$ on the vectors $e_{n}$ is given by the formulas
$$Ce_{2n}= \left[  \frac{2\nu-1}{4}\right]_{q^2}^2e_{2n},\;\;\;Ce_{2n+1}= \left[ \frac{2\nu+1}{4}\right]_{q^2}^2e_{2n+1}.$$
In the  space $\mathfrak{S}_{q\nu} $ the Casimir operator
$C$ has two  eigenvalues
$\left[\frac{2\nu \mp 1}{4}\right]_{q^2}^2$, with  eigenvectors
in the subspaces $ \mathfrak{S}_{q\nu}^\pm$
formed by the  even and odd  basis vectors $e_{n} $, respectively.
Thus  $ \mathfrak{S}_{q\nu}$ splits into the direct sum of two $su_q(1, 1)$-irreducible subspaces  $ \mathfrak{S}_{q\nu}^+$ and $ \mathfrak{S}_{q\nu}^-$.\\
In particular Macfarlane in \cite{Macfarlane} has explored the links between the $q$-Deformed Calogero-Vasiliev Oscillator
and the $q$-analogue of the parabose oscillator  of order $p = 2\nu +1$ studied in \cite{Floreanini}.
\section{Realization of the  $q$-deformed Calogero-Vasiliev oscillator}
In this section  we discuss an explicit realization of one-dimensional $q$-deformed Calogero-Vasiliev oscillator algebra.
We give an explicit expression of the representation operators $b$, $b^+$ and $N$ defined in the previous  subsection
in terms of $q$-difference  operators.
It is known that  such representation  can be realized on a Hilbert space, on
which all these operators  are supposed to be well defined and the conjugation relations in (\ref{calegero2b}) hold.\\
For this purpose we take, as Hilbert space, the space $L_{q,\alpha}^2(\R_q) $, equipped
with the scalar product
\begin{equation*}
(\!\psi_1,\psi_2\!)=\int_{-\infty}^{\infty}\psi_1(x)\overline{\psi_2(x)}|x|^{2\alpha+1}d_qx.
\end{equation*}
We, now, construct a convenient orthonormal basis  of  $ L_{q,\alpha}^2(\R_q)$ consisting of the  $(q,\alpha)$-deformed  Hermite  functions defined by
\begin{equation}\label{wave}
\phi_{n}^\alpha(x;q)=d_{n,\alpha}\sqrt{\omega_\alpha(x;q)}\tilde h_{n,\alpha}(x;q),
\end{equation}
where $\tilde h_{n,\alpha}(x;q)$, $\omega_\alpha(x;q)$ and $d_{n,\alpha}$ are given by (\ref{hnalphatild}), (\ref{omega}) and
(\ref{dncoef}),  respectively.

\begin{prop}
$\{  \phi_{n}^\alpha(x;q)   \}_{n=0}^\infty $  is a complete
orthonormal set in  $ L_{q,\alpha}^2(\R_q)$.
\end{prop}
\noindent \textbf{Proof:}\\
The  (discrete) orthogonality relation (\ref{orthohnalpht}) for $\tilde h_{n,\alpha}(x;q)$ can be written as
\begin{equation*}
\ds\int_{-\infty}^{\infty}\phi_{n}^\alpha(x;q)\phi_{m}^\alpha(x;q) |x|^{2\alpha+1}d_qx= \delta_{n,m}.
\end{equation*}
Thus $\{ \phi_{n}^\alpha(x;q) \}_{n=0}^\infty $ is an orthonormal set in  $L_{q,\alpha}^2(\R_q)$.
Let us prove that it  is complete.
Suppose that there exists $f\in L_{q,\alpha}^2(\R_q)$ orthogonal to all $ \phi_{n}^\alpha(x;q)$, that is
$$ \int_{-\infty}^{\infty} \phi_{n}^\alpha(x;q) f(x)|x|^{2\alpha +1}d_qx=0,\;\;\;\mbox{for all}\;\; n\in \N .$$
By using the inverse formula (\ref{monometild}), we obtain
$$ \int_{-\infty}^{\infty}\sqrt{\omega_\alpha(x;q)} x^n  f(x)|x|^{2\alpha +1}d_qx=0,\;\;\;\mbox{for all}\;\;  n\in \N .$$
$$
So,
\begin{array}{lll}
\mathcal{F}_D^{\alpha, q}(\sqrt{\omega_\alpha(.;q)} f )(\lambda) &=& K_{\alpha}\ds\int_{-\infty}^{+\infty}
 \sqrt{\omega_\alpha(x;q)} f(x)\psi_{-\lambda}^{\alpha, q}(x) |x|^{2\alpha +1} d_qx,\\
&=&K_{\alpha}\ds\sum_{n=0}^{\infty}b_{n}(-i\lambda;q^2)\int_{-\infty}^{\infty}\sqrt{\omega_\alpha(x;q)} x^n  f(x)|x|^{2\alpha +1}d_qx\\
&=&0.
\end{array}
$$
But, since $f\in L_{q,\alpha}^2(\R_q)$ and $\omega_\alpha(.;q)$ is bounded on $\R_q$, we deduce that $\sqrt{\omega_\alpha(.;q)}f\in L_{q,\alpha}^2(\R_q)$ and from the Plancheral theorem,
 we get $f=0$. \ef

We denote by $\delta_{q}$   the $q$-dilatation operator in the variable $x$, defined by
$\delta_{q}f(x)=f(qx)$, and the operator of multiplication by a
function $g$ will be denoted also by $g$.\\

Let $\mathfrak{S}_{q\alpha}$ be the  finite linear  span of $(q,\alpha)$-deformed  Hermite  functions $\phi_{n}^\alpha(x;q)$.
From  the forward and backward shift operators (\ref{forwardshifttild}) and  (\ref{backward}),
we define the  operators $a$ and $a^+$  on  $\mathfrak{S}_{q\alpha}$   in a $2\times 2$ matrix form  by
\begin{equation}\label{oper-a}
af=\frac{q^\frac{1}{2}}{\sqrt{1-q}x}
\begin{pmatrix}
\delta_{q^{-1}}\sqrt{1+q^{-2\alpha-1}x^2}-1 & 0 \\
0 &  \delta_{q^{-1}}\sqrt{1+q^{-2\alpha-1}x^2}- q^{2\alpha+1}
\end{pmatrix}%
\begin{pmatrix}
f_e \\
f_o
\end{pmatrix}%
\end{equation}
\begin{equation}\label{oper-a+}
a^{+}f=\frac{q^{2\alpha+\frac{3}{2}}}{\sqrt{1-q}x}
\begin{pmatrix}
\sqrt{1+q^{-2\alpha-1}x^2}\delta_{q}-1 & 0 \\
0 &  \sqrt{1+q^{-2\alpha-1}x^2}\delta_{q}- q^{-2\alpha-1}
\end{pmatrix}%
\begin{pmatrix}
f_e \\
f_o
\end{pmatrix}%
\end{equation}
where $f_e$ and $f_o$ are respectively the even and the odd parts of $f\in\mathfrak{S}_{q\alpha}$.\\
The reader may  verify that these operators are indeed mutually adjoint
in the Hilbert space $ L_{q,\alpha}^2(\R_q)$.\\
The  action of the operators  $a$ and $a^+$ on the
basis   $\{\phi_{n}^\alpha(x;q)\}_{n=0}^\infty $ of   $ L_{q,\alpha}^2(\R_q)$ leads to the explicit results:
\begin{prop}
\begin{eqnarray}
a \phi_{0}^\alpha(x;q) &=&0,\label{aphi0}\\
a \phi_{n}^\alpha(x;q) &=&\sqrt{\left \llbracket n \right\rrbracket_{q,\alpha}}\phi_{n-1}^\alpha(x;q),\;\;\;n\ge1,\label{aphin}\\
a^+\phi_{n}^\alpha(x;q) &=&\sqrt{\left\llbracket n+1\right\rrbracket_{q,\alpha}}\phi_{n+1}^\alpha(x;q),\label{a+phin}\\
\phi_{n}^\alpha(x;q)&=& (n!_{q,\alpha})^{-\frac{1}{2}}a^{+n} \phi_{0}^\alpha(x;q)\label{phinexp},
\end{eqnarray}
where  $\left\llbracket n \right\rrbracket_{q,\alpha}$ is defined by (\ref{nfactalpha}).
\end{prop}
\noindent \textbf{Proof:}\\
(\ref{aphi0}) is an immediate consequence of the definition (\ref{wave}).
(\ref{aphin}) and (\ref{a+phin}) follow from the forward and backward shift operators (\ref{forwardshifttild}) and  (\ref{backward}) and from the fact that
\begin{equation*}
d_{n,\alpha}=\frac{q^{n-\frac{1}{2}}\sqrt{\left\llbracket n \right\rrbracket_{q,\alpha}}}{\sqrt{1-q}\left\llbracket n\right\rrbracket_q}d_{n-1,\alpha}.
 \end{equation*}
(\ref{phinexp}) is a consequence of (\ref{a+phin}).
\ef

From  (\ref{aphin}) and (\ref{a+phin}),  one deduces that
\begin{equation}\label{eqaa+}
a^+a\phi_{n}^{\alpha}(x;q) =\left\llbracket n\right\rrbracket_{q,\alpha}\phi_{n}^{\alpha}(x;q)
\end{equation}
and
\begin{equation}\label{eqa+a}
 aa^+\phi_{n}^{\alpha}(x;q) =\left\llbracket n+1\right\rrbracket_{q,\alpha}\phi_{n}^{\alpha}(x;q).
\end{equation}

The number operator $N$ is defined in this case by the relations
\begin{equation}\label{op-numb}
a^+ a=\left\llbracket N\right \rrbracket_{q,\alpha},\;\;\;aa^+=\left\llbracket N+1\right\rrbracket_{q,\alpha}\;\;\;\mbox{on}\;\;\mathfrak{S}_{q\alpha}.
\end{equation}
The formulas  (\ref{op-numb}) can be inverted to  determine an explicit expression  of the  operator $N$  as follows
\begin{equation}\label{particule-operator}
N:=\frac{1}{2\log q}\log\left[1-(1-q)aa^+\right]+\frac{1}{2\log q}\log\left[1-(1-q)a^+a\right]-\alpha-1.
\end{equation}
From (\ref{eqaa+}), (\ref{eqa+a}) and  (\ref{particule-operator}), we obtain
\begin{equation}
 N\phi_{n}^{\alpha}(x;q)=n\phi_{n}^{\alpha}(x;q)
\end{equation}
and
\begin{equation}
[N,a]=-a,\;\;\; [N,a^+]=a^+\;\;\;\mbox{on}\;\;\mathfrak{S}_{q\alpha}.
\end{equation}

Now, we shall construct explicitly the generators $b$ and $b^+$  of the $q^\frac{1}{2}$-deformed Calogero-Vasiliev algebra defined in the previous subsection by means of the operators $a$ and $a^+$  in the following way

$$ b=q^{-\frac{N+(K+1)(\alpha+\frac{1}{2})}{4}} a,\;\;\;    b^+=a^+ q^{-\frac{N+(K+1)(\alpha+\frac{1}{2})}{4}},\;\; K=(-1)^N.$$
Using  the relation
\begin {equation*}\label{relatqnumb}
[x]_{q^\frac{1}{2}}=q^{-\frac{x-1}{2}}\llbracket x\rrbracket_{q},
\end {equation*}
one easily verifies that the actions of operators $b$ and  $b^+$ on the basis $\{\phi_{n}^\nu(x;q)\}_{n=0}^\infty  $ are given by

\begin{equation}\label{bb+actions}
\begin{array}{llll}
b \phi_{0}^\alpha(x;q) &=&0,&\\
 b\phi_{n}^\alpha(x;q) &=& \sqrt{\left[ n\right]_{q^{\frac{1}{2}},\alpha}}\phi_{n-1}^\alpha(x;q)  ,&n\ge1,\\
b^+\phi_{n}^\alpha(x;q)&=& \sqrt{\left[n+1\right]_{q^{\frac{1}{2}},\alpha}} \phi_{n+1}^\alpha(x;q),\\
\end{array}
\end{equation}
where
$$\left[ 2n\right]_{q^{\frac{1}{2}},\alpha}= \left[ 2n\right]_{q^{\frac{1}{2}}}\quad\hbox{and}\quad\left[ 2n+1\right]_{q^{\frac{1}{2}},\alpha}=\left[ 2n+2\alpha+2\right]_{q^{\frac{1}{2}}}.$$
From (\ref{bb+actions}),  the   basis vectors $\phi_{n}^\alpha(x;q)$  may also be expressed in terms  of the operator $b^+$ and $\phi_{0}^\alpha(x;q)$ as follows
%
$$\phi_{n}^\alpha(x;q)=\frac{1}{\sqrt{[n]!_{q^{\frac{1}{2}},\alpha}}}(b^+)^n \phi_{0}^\alpha(x;q) ,$$
where 
$$[n]!_{q^{\frac{1}{2}},\alpha}=\left[1 \right]_{q^{\frac{1}{2}},\alpha}\left[2 \right]_{q^{\frac{1}{2}},\alpha}...\left[ n \right]_{q^{\frac{1}{2}},\alpha}.$$
From the above facts, we may check that equation (\ref{calegero2b}) holds and
\begin{equation}\label{sombb}
bb^+= \left[  N+1+\nu(1+K)\right]_{q^{\frac{1}{2}}}, \;\;\;
b^+b= \left[  N+\nu(1-K)\right]_{q^{\frac{1}{2}}},\;\;\nu=\alpha+\frac{1}{2}\;\;\mbox{on}\;\;\mathfrak{S}_{q\alpha}.
\end{equation}
We deduce from (\ref{sombb}) that  the operators $b$ and $b^+$  satisfy the relations
\begin{equation}\label{commut}
bb^+ - q^{\pm \frac{1+2\nu K}{2}}b^+b= \left[ 1+2\nu K \right]_{q^{\frac{1}{2}}}q^{\mp \frac{N+\nu -\nu K}{2}}
\;\;\;\mbox{on}\;\;\mathfrak{S}_{q\alpha}.
\end{equation}
This leads to an explicit expressions for the generators  $\{ b,\; b^+,\;N\}$   of the  $q$-deformed Calogero-Vasiliev Oscillator algebra. The corresponding
 Hamiltonian is defined from $b$ and  $b^+$ according to
\begin{equation}\label{hamilton2}
H= \frac{1}{2}\left\{b,b^+\right\}.
\end{equation}
The functions $\phi_{n}^{\alpha}(x;q)$ are eigenfunctions of this Hamiltonian with  eigenvalues
$$E_{q\alpha}(n)=\frac{1}{2}\left( \left[ n\right]_{q^{\frac{1}{2}},\alpha}+ \left[ n+1\right]_{q^{\frac{1}{2}},\alpha} \right).$$
 Thus, we recover in the limit $q\rightarrow 1$ the eigenvalues of  the Hamiltonian of the  Calogero-Vasiliev oscillator.\\

In the same manner, as in the case of $su(1, 1)$, by virtue of the results of the previous subsection,  we  construct an explicit realization of the operators $B_-$, $B_+$ and $B_0$   generators of the quantum algebra $su_{q^{\frac{1}{2}}}(1,1)$ in terms of  the oscillation operators $b$, $b^+$ and $N$ by setting

$$B_+=\gamma (b^+)^2,\;\;\; B_-=\gamma b^2,\;\;\;B_0=\frac{1}{2}(N+\alpha+1),\;\;\;\gamma=(\left[  2\right]_{q^{\frac{1}{2}}})^{-1}.$$
From (\ref{bb+actions}), we derive  the actions of these operators on the  basis $\{  \phi_{n}^\nu(x;q)   \}_{n=0}^\infty $
\begin{equation}\label{B0B+B}
\begin{array}{llll}
B_0\phi_{n}^\alpha(x;q)&=& \frac{1}{2}(n+\alpha+1)\phi_{n}^\alpha(x;q),\\
B_+\phi_{n}^\alpha(x;q)&=& \gamma\sqrt{\left[n+2\right]_{q^{\frac{1}{2}},\alpha}  \left[ n+1\right]_{q^{\frac{1}{2}},\alpha}}\phi_{n+2}^\alpha(x;q),\\
B_-\phi_{n}^\alpha(x;q) &=& \gamma\sqrt{\left[ n\right]_{q^{\frac{1}{2}},\alpha}\left[ n-1\right]_{q^{\frac{1}{2}},\alpha}}\phi_{n-2}^\alpha(x;q),\;\;n\ge1.\\
\end{array}
\end{equation}
It follows that
\begin{equation}\label{BB+act}
\begin{array}{llll}
B_-B_+\phi_{2n}^\alpha(x;q)&=& \gamma^2\left[ 2n+2\right]_{q^{\frac{1}{2}}}  \left[ 2n+2\alpha+2\right]_{q^{\frac{1}{2}}}\phi_{2n}^\alpha(x;q),\\
B_-B_+\phi_{2n+1}^\alpha(x;q)&=& \gamma^2\left[2n+2\right]_{q^{\frac{1}{2}}}\left[ 2n+2\alpha+4\right]_{q^{\frac{1}{2}}}\phi_{2n+1}^\alpha(x;q),\\
B_+B_-\phi_{2n}^\alpha(x;q) &=& \gamma^2\left[2n\right]_{q^{\frac{1}{2}}}\left[ 2n+2\alpha\right]_{q^{\frac{1}{2}}}\phi_{2n}^\alpha(x;q),\\
B_+B_- \phi_{2n+1}^\alpha(x;q)&=& \gamma^2\left[2n\right]_{q^{\frac{1}{2}}}\left[  2n+2\alpha+2\right]_{q^{\frac{1}{2}}}\phi_{2n+1}^\alpha(x;q).\\
\end{array}
\end{equation}
Using the following identity (see \cite{Biedenharn} p.58)
\begin{equation}\label{q-numb-add}
 \left[ x \right]_{q}\left[ y-z \right]_{q}
+\left[ y \right]_{q}\left[ z-x \right]_{q}
+\left[ z \right]_{q}\left[ x-y \right]_{q}=0,
\end{equation}
with $x=2n+2$, $y=-2n-2\alpha$, $z=2$ and with $x=2n+2$, $y=-2n-2\alpha-2$, $z=2$ respectively,  we obtain
$$\left[2n+2\right]_{q^{\frac{1}{2}}}  \left[ 2n+2\alpha+2\right]_{q^{\frac{1}{2}}}-\left[ 2n\right]_{q^{\frac{1}{2}}}\left[  2n+2\alpha\right]_{q^{\frac{1}{2}}}= \left[ 2\right]_{q^{\frac{1}{2}}}
\left[  4n+2\alpha+2\right]_{q^{\frac{1}{2}}},
 $$
$$
\left[ 2n+2\right]_{q^{\frac{1}{2}}}\left[ 2n+2\alpha+4\right]_{q^{\frac{1}{2}}}-
\left[ 2n\right]_{q^{\frac{1}{2}}}\left[  2n+2\alpha+2\right]_{q^{\frac{1}{2}}}=\left[  2\right]_{q^{\frac{1}{2}}}
\left[  4n+2\alpha+4\right]_{q^{\frac{1}{2}}}.
$$
By the identity $\left[  2x\right]_{q^{\frac{1}{2}}}= \left[  2\right]_{q^{\frac{1}{2}}}\left[  x\right]_{q}$, we obtain
$$ \left[  4n+2\alpha+2\right]_{q^{\frac{1}{2}}}=\left[  2\right]_{q^{\frac{1}{2}}}\left[  2n+\alpha+1\right]_{q},$$
$$ \left[  4n+2\alpha+4\right]_{q^{\frac{1}{2}}}=\left[  2\right]_{q^{\frac{1}{2}}}
\left[  2n+\alpha+2\right]_{q},
$$
from which follows the following commutation relations 
$$[B_0,\pm B]=\pm B_{\pm},\;\;\;[B_-, B_+]=\left[ 2B_0 \right]_{q}\;\;\;\mbox{on}\;\;\mathfrak{S}_{q\alpha}$$
and the conjugation relations
$$B_0^*=B_0,\;\;\; B_+^*=B_- \;\;\;\mbox{on}\;\;\mathfrak{S}_{q\alpha}.$$
We conclude  an explicit  realization of generators  $B_0$, $B_-$ and $B_+$  of the quantum algebra $su_{q^{\frac{1}{2}}}(1, 1)$.\\
To analyze the irreducible representations  of $su_{q^{\frac{1}{2}}}(1, 1)$  algebra,   we need
the invariant Casimir operator $C$,   which in this case has the  form:
$$ C= \left[ B_0-\frac{1}{2}\right]_{q}^2-B_+B_-.$$
From (\ref{B0B+B}) and (\ref{BB+act}) we obtain
the action  of this  operator on the  basis $\{  \phi_{n}^\alpha(x;q)   \}_{n=0}^\infty $
$$C\phi_{2n}^\alpha(x;q)=\left(\left[ n +\frac{\alpha}{2}\right]_{q}^2-\left[ n\right]_{q}\left[ n+\alpha\right]_{q}\right) \phi_{2n}^\alpha(x;q),$$
$$C\phi_{2n+1}^\alpha(x;q)=\left(\left[ n+\frac{\alpha+1}{2}\right]_{q}^2-\left[ n\right]_{q}\left[ n+\alpha+1\right]_{q}\right) \phi_{2n+1}^\alpha(x;q).$$
Using (\ref{q-numb-add}) with $x= n+\frac{\alpha}{2}$, $y=n$, $z=-\frac{\alpha}{2}$  and with
$x= n+\frac{\alpha+1}{2}$, $y=n$, $z=-\frac{\alpha+1}{2}$ respectively, we deduce
$$\left[ n+\frac{\alpha}{2}\right]_{q}^2-\left[ n\right]_{q}\left[ n+\alpha\right]_{q}=\left[\frac{\alpha}{2}\right]_{q}^2 ,$$
$$\left[ n+\frac{\alpha+1}{2}\right]_{q}^2-\left[ n\right]_{q}\left[ n+\alpha+1\right]_{q}=\left[\frac{ \alpha +1}{2}\right]_{q}^2 .$$
Then, the Casimir operator $C$ has two  eigenvalues
$ \left[\frac{2\alpha+1  \mp 1}{4}\right]_{q}^2 $
in the subspaces $\mathfrak{S}_{q\alpha}^\pm$
formed by the  even and odd  basis vectors $\{  \phi_{n}^\alpha(x;q)   \}_{n=0}^\infty $, respectively.
Thus  $ \mathfrak{S}_{q\alpha}$ splits into the direct sum of two $su_{q^{\frac{1}{2}}}(1, 1)$-irreducible subspaces  $ \mathfrak{S}_{q\alpha}^+$ and $\mathfrak{S}_{q\alpha}^-$.\\
In particular Macfarlane  in \cite{Macfarlane} showed that  this oscillator realises the $q$-deformed parabose oscillator  of order $p = 2\nu +1$  studied in \cite{Floreanini}.

Hence we derive an explicit realizations of the annihilation and creation operators of  $q$-deformed parabose oscillator in terms of $q$-difference operators.
\subsection{The $q$-coherent states }

The  normalized $q$-coherent state   $ \varphi_\zeta(x;q)$  related to  the  $q$-deformed Calogero-Vasiliev oscillator
 is  defined as  the  eigenfunction of the annihilation operator $a$ with eigenvalue $\zeta\in \C$,
\begin{equation}\label{aphi}
 a\varphi_\zeta(x;q)=\zeta\varphi_\zeta(x;q)\;\;\mbox{on}\;\;\mathfrak{S}_{q\alpha}.
\end{equation}
\begin{thm}
The $q$-coherent states  are of the form
\begin{equation}\label{coherent}
\varphi_\zeta(x;q)=\ds\frac{c_{\alpha}\sqrt{\omega_\alpha(x;q)}}{\sqrt{e_{q,\alpha}(-(1-q)\zeta^2)}}e_{q^2}(-q(1-q)\zeta^2)E_{q,\alpha}(q^{\frac{1}{2}}(1-q)^{\frac{1}{2}}x\zeta),
\end{equation}
where $ c_{\alpha}$ is given in (\ref{dncoef}).
\end{thm}
\noindent \textbf{Proof:}\\
By expressing $\varphi_\zeta(x;q) $ in terms of the wave functions $\phi_{n}^\alpha(x;q)$, we get
\begin{equation}\label{phiexpan}
\varphi_\zeta(x;q)=\ds\sum_{n=0}^{+\infty}f_{n,\alpha}(q)\phi_{n}^\alpha(x;q).
\end{equation}
From the eigenvalue equations (\ref{aphi0}) and (\ref{aphin}),  we can write
\begin{equation}\label{aphiexpan}
 a\varphi_\zeta(x;q)=\ds\sum_{n=1}^{+\infty}f_{n,\alpha}(q) \sqrt{\llbracket n\rrbracket_{q,\alpha}} \phi_{n-1}^\alpha(x;q).
\end{equation}
Replace $\varphi_\zeta(x;q)$ by the series (\ref{phiexpan}) in  (\ref{aphi})  and equate the coefficients of $ \phi_{n}^\alpha(x;q) $ on both sides to get
$$f_{n+1,\alpha}(q) \sqrt{\llbracket n+1\rrbracket_{q,\alpha}} =\zeta f_{n,\alpha}(q).$$
By iterating  the last relation, we get
since $  f_{0,\alpha}(q)= C_0=C_0(\zeta) $, the relations
$$
f_{1,\alpha}(q)=\frac{C_0\zeta}{\sqrt{\llbracket 1\rrbracket_{q,\alpha}} },\;\;\;
f_{2,\alpha}(q)=\frac{C_0\zeta^2}{\sqrt{2!_{q,\alpha} }},\;\;\;...\;\;\;,
f_{n,\alpha}(q)=\frac{C_0\zeta^n}{\sqrt{n!_{q,\alpha} }},$$
which, inserted into the expansion (\ref{phiexpan}), give 
$$\varphi_\zeta(x;q)=C_0(\zeta)\ds\sum_{n=0}^{+\infty}\frac{\zeta^n}{\sqrt{n!_{q,\alpha} }}\phi_{n}^\alpha(x;q).$$
Now, for $\zeta, \zeta'\in \C$, we have the scalar product
\begin{eqnarray*} \ds&&\int_{-\infty}^{+\infty}\varphi_{\zeta}(x;q)\varphi_{\zeta'}(x;q)|x|^{2\alpha+1}d_qx\\ &=& C_0(\zeta)C_0(\zeta')
\ds\sum_{n,k=0}^{+\infty}\frac{\zeta^n{\zeta'}^k}{\sqrt{n!_{q,\alpha} }\sqrt{k!_{q,\alpha} }}\ds\int_{-\infty}^{+\infty}\phi_{n}^\alpha(x;q)\phi_{k}^\alpha(x;q)|x|^{2\alpha+1}d_qx.\end{eqnarray*}
But, the orthogonality relation (\ref{orthohnalpht} ) implies that
$$\ds\int_{-\infty}^{+\infty}\varphi_{\zeta}(x;q)\varphi_{\zeta'}(x;q)|x|^{2\alpha+1}d_qx= C_0(\zeta)C_0(\zeta')\ds\sum_{n=0}^{+\infty}\frac{\zeta^n{\zeta'}^n}{n!_{q,\alpha} }. $$
By the relation (\ref{q-expo-alpha}), we get
$$\ds\int_{-\infty}^{+\infty}\varphi_{\zeta}(x;q)\varphi_{\zeta'}(x;q)|x|^{2\alpha+1}d_qx= C_0(\zeta)C_0(\zeta')e_{q,\alpha}(-(1-q)\zeta\zeta')   . $$
The normalized condition requires to choose $C_0(\zeta)=\ds\frac{1}{\sqrt{e_{q,\alpha}(-(1-q)\zeta^2)}}$.\\
So, we can write
$$ \varphi_\zeta(x;q)=\ds\frac{1}{\sqrt{e_{q,\alpha}(-(1-q)\zeta^2)}}\ds\sum_{n=0}^{+\infty}\frac{\zeta^{n}}{\sqrt{n!_{q,\alpha} }}\phi_{n}^\alpha(x;q).$$
From the relations   (\ref{wave} ) and (\ref{dncoef}), we obtain
$$ \varphi_\zeta(x;q)= \ds\frac{\sqrt{\omega_\alpha(x;q)}}{\sqrt{e_{q,\alpha}(-(1-q)\zeta^2)}}\sum_{n=0}^{+\infty}\frac{\zeta^{n}}{\sqrt{n!_{q,\alpha} }}    c_{\alpha}q^{\frac{n^2}{2}}\frac{\left(q;q\right)_{n,\alpha}^{\frac{1}{2}}}{\left(q;q\right)_{n}}\tilde h_{n,\alpha}(x;q),$$
which can be rewritten as
$$\varphi_\zeta(x;q)=  \ds\frac{c_{\alpha}\sqrt{\omega_\alpha(x;q)}}{\sqrt{e_{q,\alpha}(-(1-q)\zeta^2)}}\sum_{n=0}^{+\infty}\frac{q^{\frac{n(n-1)}{2}}\left(q^{\frac{1}{2}}(1-q)^{\frac{1}{2}}\zeta\right)^n}{\left(q;q\right)_{n}}\tilde h_{n,\alpha}(x;q).$$
So, from  the generating function (\ref{gghntild} ) for the polynomials $\tilde h_{n,\alpha}(x;q) $,  we get
the explicit form of the normalized $q$-coherent state (\ref{coherent}).\ef

\subsection{Limit to the   Calogero oscillator }
\begin{lem}
\begin{equation}
\ds\lim_{q\rightarrow1^-}(1-q^2)^{\frac{\alpha+1}{2}}\phi_{n}^{\alpha}(\sqrt{1-q^2}x;q)=\phi_n^{\alpha+\frac{1}{2}}(x),
\end{equation}
where $\phi_n^{\mu} $ is  the  Rosenblum's Hermite function defined  by formula (\ref{defphimu}).
\end{lem}
\noindent \textbf{Proof:}\\
We have
\begin{equation*}
\ds\lim_{q\rightarrow1^-}\frac{\tilde h_{n,\alpha}(\sqrt{1-q^2}x;q)}{(1-q^2)^{\frac{n}{2}}}=\frac{H_n^{\alpha+\frac{1}{2}}(x)}{2^n},
\end{equation*}
where $ H_n^{\alpha+\frac{1}{2}}(x)$ is  the  Rosenblum's Hermite polynomials.
\begin{equation}\label{limomega}
\ds\lim_{q\rightarrow1^-} \omega_\alpha(\sqrt{1-q^2}x;q) =exp(-x^2).
\end{equation}
\begin{equation*}
\begin{array}{lll}
\ds\lim_{q\rightarrow1^-}d_{n,\alpha}(1-q^2)^{\frac{n}{2}}&=& \ds\lim_{q\rightarrow1^-}c_{\alpha}\ds\lim_{q\rightarrow1^-}
q^{\frac{n^2}{2}}\frac{\left(q;q\right)_{n,\alpha}^{\frac{1}{2}}}{\left(q;q\right)_{n}}(1-q^2)^{\frac{n}{2}}\\
&=& 2^{\frac{n}{2}}\frac{\sqrt{\gamma_{\alpha+\frac{1}{2}}(n)}}{n!}\ds\lim_{q\rightarrow1^-}c_{\alpha}.
\end{array}
\end{equation*}
We have the limits (see \cite{andrews} Theorem 10.2.4)
\begin{equation*}
\begin{array}{lll}
\ds\lim_{q\rightarrow1^-}\frac{(-q^{-2\alpha-1};q^2)_\infty}{(-q;q^2)_\infty}&=&2^{\alpha+1}\\
\ds\lim_{q\rightarrow1^-}\frac{(-q^{2\alpha+3};q^2)_\infty}{(-q;q^2)_\infty}&=&2^{-\alpha-1}\\
\ds\lim_{q\rightarrow1^-}\frac{(q^{2\alpha+2};q^2)_\infty(1-q^2)^\alpha}{(q^2;q^2)_\infty}&=& \ds\lim_{q\rightarrow1^-}\frac{1}{\Gamma_{q^2}(\alpha+1)}=\frac{1}{\Gamma(\alpha+1)}.
\end{array}
\end{equation*}
Then, 
$$\ds\lim_{q\rightarrow1^-} (1-q^2)^{\frac{\alpha+1}{2}}c_{\alpha}=\ds\frac{1}{\sqrt{\Gamma(\alpha+1)}}$$
\begin{equation}\label{limitdnhn2}
\ds\lim_{q\rightarrow1^-} (1-q^2)^{\frac{\alpha+1}{2}}d_{n,\alpha}\tilde h_{n,\alpha}(\sqrt{1-q^2}x;q)
=\left(\frac{\gamma_{\alpha+\frac{1}{2}}(n)}{\Gamma(\alpha+1)}\right)^{\frac{1}{2}}\frac{ H_n^{\alpha+\frac{1}{2}}(x)  }{2^{\frac{n}{2}}n!}
\end{equation}
\begin{equation*}
\begin{array}{lll}
\ds\lim_{q\rightarrow1^-}(1-q^2)^{\frac{\alpha+1}{2}} \phi_{n}^{\alpha}(\sqrt{1-q^2}x;q)&=&
\left(\frac{\gamma_{\alpha+\frac{1}{2}}(n)}{\Gamma(\alpha+1)}\right)^{\frac{1}{2}}exp(-\frac{x^2}{2})\frac{ H_n^{\alpha+\frac{1}{2}}(x)}{2^{\frac{n}{2}}n!}\\
&=&\phi_{n}^{\alpha+\frac{1}{2}}(x).
\end{array}
\end{equation*}\ef
In the limit as $q\rightarrow  1^-$  the $q$-Calogero-Vasiliev oscillator reduces to the Calogero oscillator. To show this, one first verifies easily that
\begin{equation*}
a\phi_{n}^\alpha(x;q)=\sqrt{q(1-q)\omega_\alpha(x;q)}\Delta_{\alpha,q}^+ \left(d_{n,\alpha}\tilde  h_{n,\alpha}(.;q)\right)(x),
\end{equation*}
where $\Delta_{\alpha,q}^+ $ is given by (\ref{defdeltaalphaplus}).
One rescales  $x\rightarrow \sqrt{1-q^2}x$, we get
\begin{equation*}
a\phi_{n}^\alpha(\sqrt{1-q^2}x;q)=\frac{\sqrt{q\omega_\alpha(\sqrt{1-q^2}x;q)}}{\sqrt{1+q}}\Delta_{\alpha,q}^+ \left(d_{n,\alpha}\tilde  h_{n,\alpha}(\sqrt{1-q^2}x;q)\right).
\end{equation*}
Using the limits (\ref{limomega}), (\ref{limitdnhn2} ) and (\ref{limdunkl}), we find that
\begin{equation*}
\lim_{q\rightarrow 1^-} (1-q^2)^{\frac{\alpha+1}{2}} a\phi_{n}^\alpha(\sqrt{1-q^2}x;q)=
\frac{\exp(-\frac{x^2}{2})}{\sqrt2}\Lambda_{\alpha+\frac{1}{2}}
\left(\frac{\gamma_{\alpha+\frac{1}{2}}^{\frac{1}{2}}(n)    }{\Gamma^{\frac{1}{2}}(\alpha+1)2^{\frac{n}{2}}n!}H_n^{\alpha+\frac{1}{2}}(x) \right).
\end{equation*}
By definition of  the  Rosenblum's Hermite function $\phi_n^\mu $  (\ref{defphimu}) and the properties of the Dunkl  operator $\Lambda_{\alpha}$, we have
\begin{equation*}
\lim_{q\rightarrow 1^-} (1-q^2)^{\frac{\alpha+1}{2}} a\phi_{n}^\alpha(\sqrt{1-q^2}x;q)=\frac{1}{\sqrt2}(\Lambda_{\alpha+\frac{1}{2}} +xI)\phi_n^{\alpha+\frac{1}{2}}(x),
\end{equation*}
where $I$ is the identity operator. In the same way, we can write
\begin{equation*}
a^+\phi_{n}^\alpha(x;q)=\sqrt{q(1-q)\omega_\alpha(x;q)}d_{n,\alpha}\left(-H_{\alpha,q}\Delta_{\alpha,q}+
\frac{x}{1-q}\delta_q\right)\tilde  h_{n,\alpha}(x;q),
\end{equation*}
where $ \Delta_{\alpha,q}$ is the operator (\ref{defdeltaalpha}) and
\begin{equation*}H_{\alpha,  q}: f = f_e + f_o \longmapsto f_e
+ q^{2\alpha +1} f_o.
\end{equation*}
Hence, we get
\begin{equation*}
a^+\phi_{n}^\alpha(\sqrt{1-q^2}x;q)=
\sqrt{q\omega_\alpha(\sqrt{1-q^2}x;q)}d_{n,\alpha}\left(-\frac{1}{\sqrt{1+q}}H_{\alpha,q}\Delta_{\alpha,q}+
\sqrt{1+q}x\delta_q\right)\tilde  h_{n,\alpha}(\sqrt{1-q^2}x;q).
\end{equation*}
By  (\ref{limomega}), (\ref{limitdnhn2} ) and (\ref{limdunkl}), we obtain
\begin{equation*}
\begin{array}{lll}
\lim_{q\rightarrow 1^-} (1-q^2)^{\frac{\alpha+1}{2}} a^+\phi_{n}^\alpha(\sqrt{1-q^2}x;q)&=&
\exp(-\frac{x^2}{2})\left[-\frac{1}{\sqrt2}\Lambda_{\alpha+\frac{1}{2}}+\sqrt2 xI\right]
\left(\frac{\gamma_{\alpha+\frac{1}{2}}^{\frac{1}{2}}(n)    }{\Gamma^{\frac{1}{2}}(\alpha+1)2^{\frac{n}{2}}n!}H_n^{\alpha+\frac{1}{2}}(x) \right)\\
&=& \frac{1}{\sqrt2}(-\Lambda_{\alpha+\frac{1}{2}} +xI)\phi_n^{\alpha+\frac{1}{2}}(x)\\
\end{array}
\end{equation*}
Note that if we replace $ \alpha+\frac{1}{2}$ by $\nu$ we obtain  the  annihilation and creation  operators of one-dimensional  two-particle Calogero oscillator given by (\ref{annih-crea}).


\end{document}